\documentclass{article}
\usepackage[T1]{fontenc}
\usepackage[utf8]{inputenc}
\usepackage[]{ismir} 
\usepackage{amsmath,cite,url}
\usepackage{graphicx}
\usepackage{color}

\usepackage{adjustbox}
\usepackage{multirow}
\usepackage{makecell}
\usepackage{bbding}
\usepackage{algpseudocode}
\usepackage{threeparttable,booktabs}
\usepackage{amssymb}
\usepackage{stmaryrd} 
\usepackage{todonotes}

\title{MPEcho: A Melody and Phoneme-Aware Generative Framework for Controllable Cover Song Generation}

\multauthor
  {Wei-Jaw Lee$^{1, 2, 3}$ \hspace{1cm} Hsuan-Yu Yeh$^{1, 3}$ \hspace{1cm} Ting-Yi Hu$^{1, 2}$}
  {{\bf Chih-Pin Tan$^{1, 2}$ \hspace{1cm} Fang-Duo Tsai$^{1, 3}$ \hspace{1cm} Yi-Hsuan Yang$^{1, 2, 3}$}\\
  $^1$\,GICE, National Taiwan University; $^2$\,AI-CoRE, National Taiwan University; $^3$\,Taiwan AI Labs\\
  {\tt\small weijaw2000@gmail.com, yhyangtw@ntu.edu.tw}
  }

\def\authorname{Wei-Jaw Lee, Hsuan-Yu Yeh, Ting-Yi Hu, Chih-Pin Tan, Fang-Duo Tsai and Yi-Hsuan Yang}

\begin{document}

\maketitle

\begin{abstract}
Cover song generation (CSG) should preserve the melodic and linguistic content of a reference song while recreating the remaining musical components. The state-of-the-art model SongEcho utilizes $F_0$ sequences and voiced/unvoiced (V/UV) tags for conditioning; however, implicit linguistic information from V/UV tags cannot guarantee lyric accuracy, leading to a high phoneme error rate (PER). Inspired by singing voice synthesis (SVS), we propose MPEcho, which integrates a phoneme encoder and a length regulator (LR) into the SongEcho framework. By providing explicit phoneme-level conditioning and precise temporal boundaries, MPEcho significantly reduces PER. To enable this, we developed Phonsa, a Whisper-based automatic transcription model that provides high-precision phoneme-level annotations for singing voices, overcoming the scarcity of high-quality audio-phoneme pairs. Experimental results validate the effectiveness of Phonsa for alignment and MPEcho for end-to-end CSG. 
The audio samples, code and weights can be accessed from \url{https://lonian6.github.io/MPEcho.github.io/}.
\end{abstract}

\section{Introduction}\label{sec:introduction}

\begin{figure}
  \centering
  \includegraphics[width=0.9\columnwidth]
  {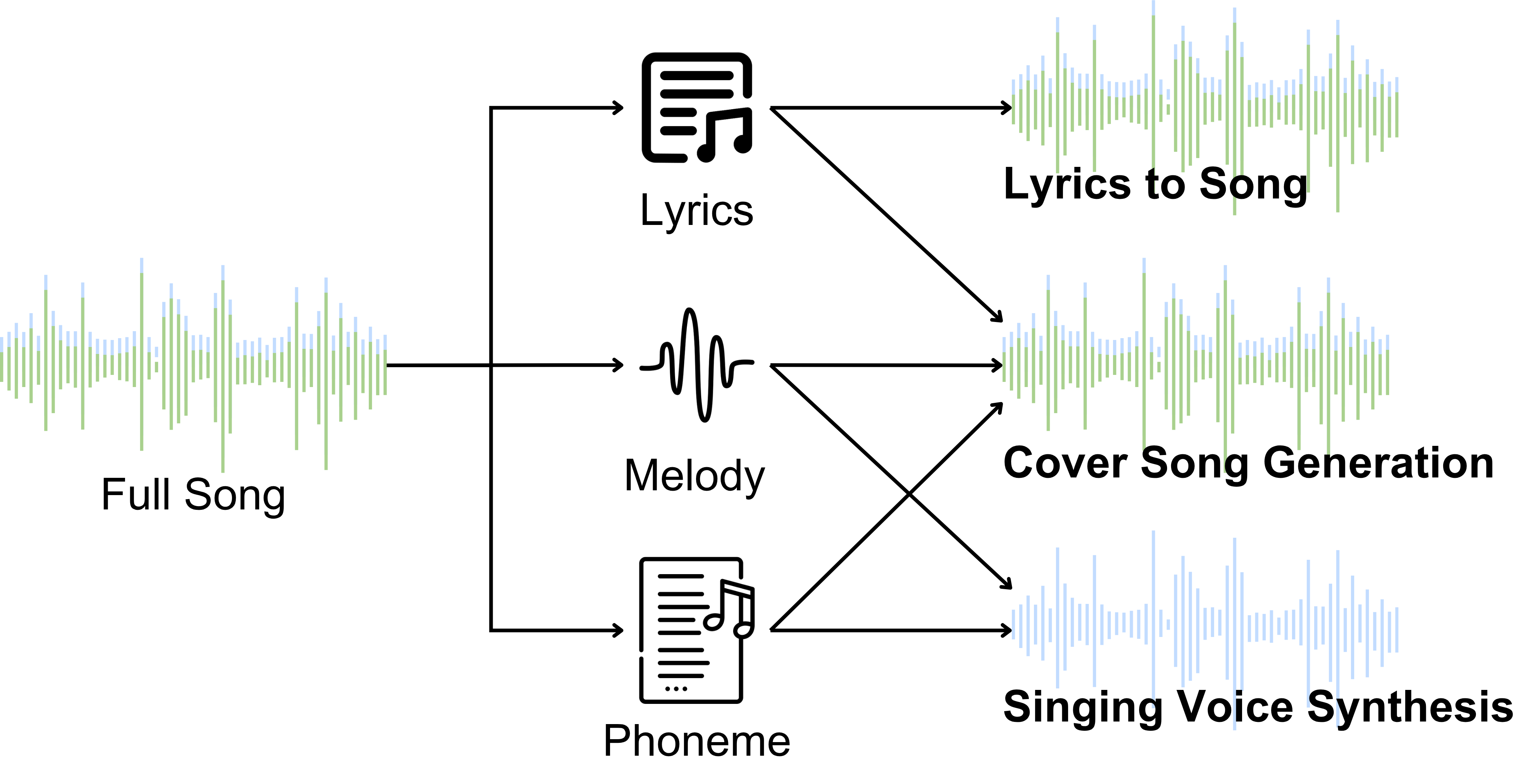}
  \caption{Conceptual comparison of lyrics-to-song generation, cover song generation, and singing voice synthesis.}
  \label{fig:task_def}
\end{figure}

Cover song generation (CSG) is an emerging generative task in music information retrieval and generative AI~\cite{li2026songecho}, requiring the preservation of a song's core melody and lyrical structure while enabling creative reinterpretation of vocal performance and arrangement. This balance between structural preservation and creative freedom makes CSG technically challenging and distinct from the related tasks of lyrics-to-song (LTS) generation and singing voice synthesis (SVS). As illustrated in Figure~\ref{fig:task_def}, general LTS models~\cite{gong2025ace, ning2025diffrhythm, liu2025jamtinyflowbasedsong, yang2025songbloom} excel at high-fidelity generation but lack fine-grained structural control. In contrast, SVS systems~\cite{liu2022diffsinger, yun2026latent, guo2025techsinger} offer precise melody and phoneme control but are limited to isolated vocals. CSG demands seamless integration of both paradigms amidst complex accompaniments.

\begin{figure*}
  \centering
  \includegraphics[width=1.8\columnwidth]
  {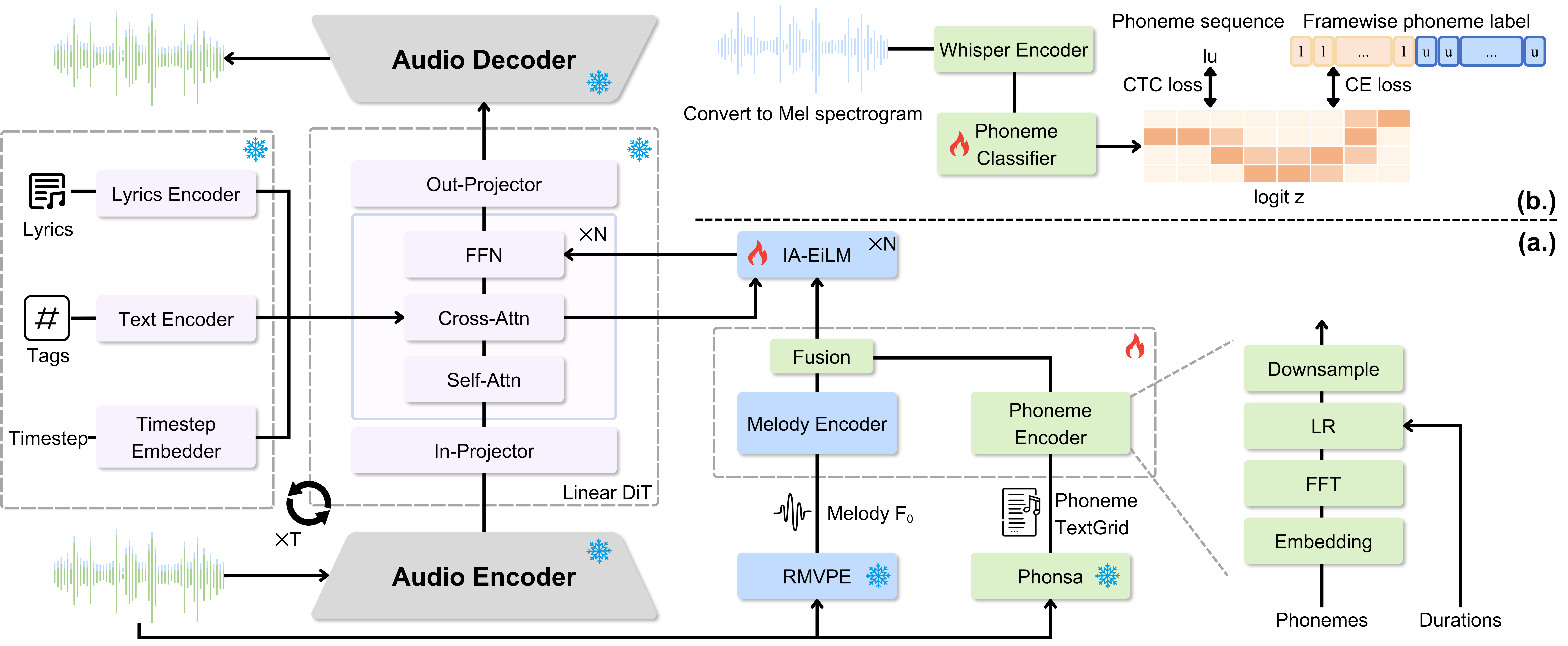}
  \caption{
  Architectural overview. (a) MPEcho incorporates structural components from ACE-Step (LTS backbone; purple) and SongEcho (melody conditioning; blue). Our contributions (green) introduce a phoneme encoder and length regulator for linguistic control. (b) The Phonsa transcription pipeline, providing the linguistic conditioning required for MPEcho.
  }
  \label{fig:model}
\end{figure*}

Building upon the lyrics-following capabilities of state-of-the-art (SOTA) open-source LTS models such as ACE-Step~\cite{gong2025ace}, Li \emph{et al.}~\cite{li2026songecho} proposed transforming an LTS model into a CSG model by incorporating melodic conditioning.  
Specifically, their model, named SongEcho, employs extracted vocal $F_0$ sequences and voiced/unvoiced (V/UV) tags to strengthen the melodic consistency between the model output and reference song. 
While representing an important step forward for CSG, our qualitative evaluation of SongEcho outputs reveals significant limitations in the linguistic consistency between the generated vocals and reference lyrics.
Neither the coarse, sentence-level lyrics control used by the backbone LTS model nor the V/UV tags offers the fine-grained phoneme-level temporal resolution requisite for accurate lyric rendering, leading to noticeable lyrical errors (45.62\% PER) that degrade perceived quality, as demonstrated on our  project page.

To address this issue, we propose MPEcho, a new model for CSG that incorporates explicit phoneme-level conditioning inspired by research on SVS. Specifically, we note that SVS models such as FastSpeech~\cite{ren2019fastspeech} employ a module named ``duration predictor'' to predict the physical temporal duration of each phoneme in the lyrics, and a module named ``length regulator'' (LR) to enforce such timing conditioning on the generative process, so that the onset and offset of each sung phoneme can be accurately controlled.  We recognize that  such precise temporal alignment of linguistic content is essential for CSG. Therefore, in addition to  melody control (`$+$melody') as implemented in SongEcho, we propose to further integrate phoneme-level control (`$+$phoneme') via an LR, as illustrated in Figure~\ref{fig:model}.
To the best of our knowledge, MPEcho represents the first end-to-end melody- and phoneme-aware CSG framework.

Unlike SVS, which requires a duration predictor to estimate timing, CSG can extract timing directly from the reference audio via an automated phonetic transcription module.
The secondary contribution of this paper is the development of Phonsa, a Whisper-based~\cite{radford2023whisper} phonetic transcription system optimized for singing voices, 
trained on SVS datasets with human-labeled phoneme timestamps. 
Phonsa provides high-precision transcription through a modified architecture incorporating chunked self-attention and specialized boundary and breath tokens. This enables the extraction of the fine-grained linguistic conditioning required by MPEcho during both training and inference.

In our experiments, we first validate the effectiveness of Phonsa for phonetic transcription using a held-out SVS dataset, demonstrating a largely improved alignment accuracy over a Montreal Forced Aligner (MFA) baseline~\cite{mcauliffe2017montreal}.
We then evaluate MPEcho for CSG through extensive ablations, showing that phoneme-level conditioning complements melody conditioning and that decomposed multi-condition guidance effectively resolves condition conflicts. 
Compared to SongEcho~\cite{li2026songecho}, MPEcho substantially reduces the PER to 18.65\%, while maintaining competitive melody consistency. MPEcho also outperforms SongEcho greatly in our subjective listening test.
Our results highlight the broader insight that SVS-derived priors can effectively enhance controllability in end-to-end full-song generation.

\section{Background}\label{sec:background}

LTS aims to synthesize full songs from lyrics and text prompts. Recent systems~\cite{dhariwal2020jukebox, gong2025ace, yang2025songbloom, ning2025diffrhythm} produce coherent long-form music with high fidelity. CSG, a subtask of music style transfer, derives new adaptations while preserving a source song's identity. While symbolic CSG (e.g., piano covers) often decouples attribute extraction from generation~\cite{choi2023pop2piano, tan2024picogen, chen2025etude}, audio-domain CSG controls generative models via signals such as the melody~\cite{tsai2025musecontrollite}. SongEcho~\cite{li2026songecho} currently represents the SOTA for full-song CSG. However, most LTS and CSG models lack fine-grained linguistic control. A notable exception is JAM~\cite{liu2025jamtinyflowbasedsong}, utilizing \emph{word}-level timing to improve LTS accuracy. 
With proposed Phonsa, we further exploit fine-grained \emph{phoneme}-level control for CSG. Our experiments will compare the effectiveness of the proposed \emph{SVS-style} phoneme-level control and the \emph{JAM-style} word-level baseline.

SVS focuses on vocal synthesis from scores and lyrics. Its fine-grained controllability relies on precise phoneme-level annotations, traditionally requiring labor-intensive manual annotation. 
Existing automatic approaches include supervised alignment (given lyrics) and unsupervised segmentation (no lyrics given). Supervised tools like MFA~\cite{mcauliffe2017montreal} often yield suboptimal results for singing due to acoustic complexities and polyphonic interference. 
Recent efforts have leveraged source separation~\cite{schulze2021phoneme} or pre-trained models~\cite{junyou2023transcription, durand2023contrastive}, yet they often prioritize word-level over phonetic-level granularity. 
An unified framework~\cite{guo2025stars} simultaneously addresses singing transcription, alignment, and refined style annotation.
Obtaining reliable phoneme timestamps is a bottleneck for bringing SVS-style precision to full-song generation, a gap we address with Phonsa.


\section{MPEcho}
\label{sec:methodology}

Figure~\ref{fig:model}(a) provides an overview of MPEcho, extending SongEcho~\cite{li2026songecho} by introducing a dedicated phoneme conditioning branch. 
It inherits SongEcho's architecture, including the ACE-Step DiT backbone~\cite{gong2025ace}, the instance-adaptive 
element-wise linear modulation (IA-EiLM) adapters, and the RMVPE~\cite{wei2023rmvpe} extractor for vocal melody conditioning.

Formally, let $\mathbf{x} \in \mathbb{R}^{T_{h} \times d}$ denote the latent audio encoded by ACE-Step's VAE. MPEcho learns a conditional denoiser $D_\theta(\mathbf{z}_t, t, c_t, c_l, c_m, c_p)$, where $\mathbf{z}_t$ is the noisy latent at timestep $t$, and $c_t$, $c_l$, $c_m$, $c_p$ denote the text prompt, lyrics, melody, and phoneme conditionings, respectively, while $c_m$ follows the processing pipeline with SongEcho; $c_p$ is a duration-expanded phoneme embedding derived from Phonsa (see Section~\ref{sec:phonsa}).
$c_m$ and $c_p$ are fused before being injected via the IA-EiLM adapters, while $c_t$ and $c_l$ are handled by ACE-Step's existing encoders. The model is trained with the flow matching objective:
\begin{equation}
\mathcal{L} = \mathbb{E}_{t,\mathbf{x},\boldsymbol{\epsilon}}
\left[\left\| D_\theta(\mathbf{z}_t, t, c_t, c_l, c_m, c_p) 
- \mathbf{x} \right\|^2 \right].
\end{equation}


\subsection{Phoneme Arrangement Condition}
\label{sec:svs-and-jam}

\begin{figure}
  \centering
  \includegraphics[width=0.9\columnwidth]
  {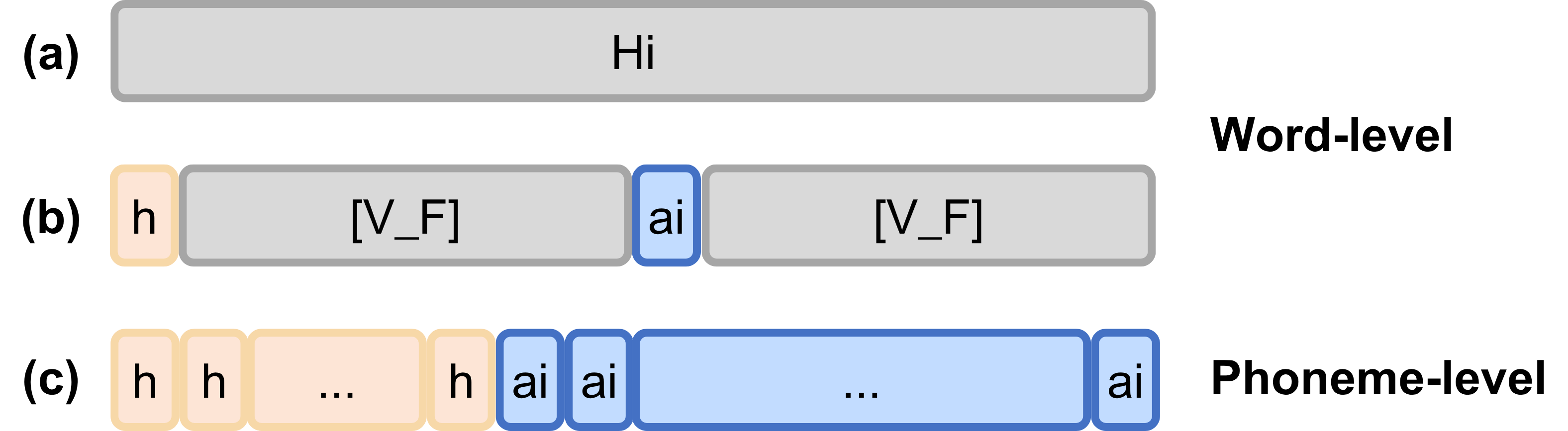}
  \caption{Comparison of phoneme arrangement methods based on word-level and phoneme-level timestamps. (a) Word arrangement for the word \texttt{Hi} based on word-level timestamps. (b) The \emph{Jam-style}~\cite{liu2025jamtinyflowbasedsong} phoneme arrangement based on word-level timestamps, incorporating \texttt{[V\_F]} (vocal filter) as a special filler token. (c) The proposed \emph{SVS-style} phoneme arrangement based on phoneme-level timestamp derived from Phonsa and LR.}
  \label{fig:phoneme_arrangement}
\end{figure}

The phonetic conditioning branch takes a phoneme sequence $\mathcal{P}=[p_1, p_2, \dots, p_n]\in \mathbb{N}^n$ and its corresponding duration sequence $\mathcal{D}=[d_1, d_2, \dots, d_n] \in \mathbb{N}^n$ as input, both derived from the onset and offset timestamps of Phonsa (see Section~\ref{sec:phonsa} for details), where $n$ denotes the number of phonemes. The tokens are mapped to embeddings and processed through a 4-layer, 2-head Feed-Forward Transformer (FFT) block, yielding hidden representations $P' \in \mathbb{R}^{n\times D_p}$ (where $D_p=256$). To align the phonetic information with the temporal axis of the audio, the length regulator (LR) expands $P'$ by repeating each phoneme embedding $p'_j$ according to its duration $d_j$. The expanded sequence $p' \in \mathbb{R}^{T_p \times D_p}$ is then passed through a downsampler and padding to ensure its temporal length $T_p$ is consistent with the DiT hidden states, yielding the phoneme condition $c_p$ defined in Section~\ref{sec:methodology}.


We refer to the sequence processed via FFT and LR as the SVS-style phoneme arrangement condition, as it preserves precise phoneme-level timing boundaries. We further investigate the impact of phonetic timing precision by comparing our approach with a word-level timing phoneme arrangement adopted from JAM \cite{liu2025jamtinyflowbasedsong}, referred to as the Jam style. 
As shown in Figure~\ref{fig:phoneme_arrangement}, the SVS-style approach provides more structured and fine-grained guidance, which is critical for singing voices where phoneme durations vary substantially due to melodic expression.

\subsection{Inference-Time Optimization}
\label{sec:cfg}
Classifier free guidance (CFG) is an important inference-time optimization, also found critical in our experiments. Given the denoiser $D_\theta$ defined in Section~\ref{sec:methodology} and omitting $z_{t}$ and $t$ for simplification, the CFG output is:
\begin{equation}
  \hat{D}_\text{CFG} = D_{\theta}(c) + (\omega-1)\Delta D_{t}\,,
\label{eq1}  
\end{equation}
where $\Delta D_{t}=D_{\theta}(c)-D_{\theta}(\phi)$ and $\omega$ is the guidance scale. With high guidance scale (i.e., $\omega$ is large), researchers observe oversaturation and artifacts in CFG. Adaptive projection guidance (APG)~\cite{sadat2024eliminating} seeks to eliminate this issue by decomposing the guidance $\Delta D_{t}$ 
via Gram-Schmidt orthogonal decomposition. The adjusted guidance $\Delta D_{t}(\eta)$ is then formulated as $\Delta D_{t}(\eta)=\Delta D^{\perp}_{t}+\eta\Delta D^{\parallel}_{t}$, where $\eta\le 1$ is a hyperparameter. When $\eta=1$, APG is equivalent to the original CFG.

SongEcho adapts APG for multi-condition control by replacing the reference basis $D_\theta(c)$ with:
\begin{equation}
 \alpha D_{\theta}(c_{t}, \phi) + (1-\alpha)D_{\theta}(c_{t}, c_{m})\,, 
 \label{eq2}
\end{equation}
with $\alpha=0.5$. However, this entangles the reference basis and reduces controllability. Following previous works~\cite{brooks2023instructpix2pix, tsai2025musecontrollite}, we instead apply multiple CFG on APG:
\begin{equation}
\begin{split}
    \hat{D}_\text{CFG}' = &D_{\theta}(\phi,\phi,c_{m},c_{p}) + 
    (\omega_{a}-1)\Delta D_{t}^{\phi,\phi,\hat{c}_{m},\hat{c}_{p}}\\
    +&D_{\theta}(\phi,c_{l},c_{m},c_{p}) + 
    (\omega_{l}-1)\Delta D_{t}^{\phi,\hat{c}_{l},c_{m},c_{p}}\\
    +&D_{\theta}(c_{t},c_{l},c_{m},c_{p}) + 
    (\omega_{t}-1)\Delta D_{t}^{\hat{c}_{t},c_{l},c_{m},c_{p}}\\
    -&D_{\theta}(\phi,c_{l},c_{m},c_{p})
    -D_{\theta}(\phi,\phi,c_{m},c_{p})\,,
\end{split}
\label{eq3}
\end{equation}
where $\omega_t$, $\omega_l$, $\omega_a$ are the guidance scales for text, lyrics, and time-varying controls (melody and phoneme), respectively, and $\Delta D_{t}^{c_{i|i\in\{t, l, m, p, \phi\}}, \hat c_{j|j\in\{t, l, m, p\}}} = D_{\theta}(c_{i}, c_{j})-D_{\theta}(c_{i}, \phi)$. We validate this approach in our experiments.

\section{Phonsa}
\label{sec:phonsa}
%

Figure \ref{fig:model}(b) offers an overview of Phonsa, a unified framework for phoneme-level forced alignment and segmentation, building upon and extending prior word-level approach~\cite{junyou2023transcription}.
Given a singing voice audio signal represented by acoustic features $x = [x_1, x_2, \dots, x_T]$, where $x_t$ denotes the Mel-spectrogram at frame $t$, our goal is to estimate both i) a frame-level phoneme tokens sequence $\rho = [\rho_1, \rho_2, \dots, \rho_T]$, where $\rho_t \in \mathcal{C}$, and ii) a sequence of phoneme segments with temporal boundaries composing of onsets $s_i$ and offsets $e_i$, $A = [(s_i, e_i)]_{i=1}^N$, where each segment corresponds to a phoneme instance and satisfies $1 \le s_i \le e_i \le T$.
From $\rho$ and $A$, we can get $\mathcal{P}$ and $\mathcal{D}$.

A neural classifier $f_\theta$ produces frame-level phoneme posterior probabilities $P \in \mathbb{R}^{T \times C}$, where $P_{t,c}$ denotes the probability of phoneme class $c$ at frame $t$. Based on $P$, we recover $\rho$ and $A$ via a monotonic decoding procedure that enforces temporal consistency.

Our framework is designed to support two settings: i) \textit{alignment}, where ground-truth phoneme or lyric sequences are available and guide the decoding process, and ii) \textit{segmentation}, where no textual supervision is provided and phoneme boundaries must be inferred solely from acoustic evidence. Formally, both tasks can be expressed as $(\rho, A) = \mathrm{Decode}(P)$ under different constraints. We focus on the alignment setting here, where ground-truth lyrics are available to constrain the decoding process.

The key improvements of Phonsa over the prior work \cite{junyou2023transcription} are three-fold. 
First, we move from word-level to phoneme-level modeling, enabling finer-grained temporal control for singing voice, where timing variations are more pronounced. 
Second, we replace the RNN-based classifier \cite{junyou2023transcription} with a chunked self-attention architecture, which improves computational efficiency through parallelization while better capturing long-range temporal dependencies in variable-length singing inputs.
Third, we introduce specialized breath and boundary tokens into the phoneme set $\mathcal{C}$ to further improve the prediction accuracy. The breath token enables the model to align and synthesize breathing sounds appropriately, while the boundary token provides a discriminative feature for the Viterbi algorithm to correctly segment consecutive identical phonemes.


The model is trained via multi-task learning with two complementary objectives $\mathcal{L} = \mathcal{L}_{\text{CTC}} + \lambda \mathcal{L}_{\text{CE}},$
where $\mathcal{L}_{\text{CTC}}$ provides sequence-level supervision for alignment, and $\mathcal{L}_{\text{CE}}$ acts as a frame-level auxiliary loss to stabilize local phoneme classification.

\section{Implementation Details}\label{sec:experimental setup}

For Phonsa, we use M4Singer \cite{zhang2022msinger}, Opencpop \cite{wang2022opencpop} for training. A subset of samples is randomly selected from each of these training datasets to serve as validation data during training. We use the GTsinger dataset \cite{zhang2024gtsinger} exclusively as our test set. The training, validation, and test sets comprised audio data totaling 30.92 hours, 1.64 hours, and 16.54 hours, respectively. All the three datasets are publicly available and are for Mandarin. 

The training of Phonsa follows the backbone model and config of the prior work~\cite{junyou2023transcription}.
We train on a single NVIDIA RTX 3090 GPU with a batch size of 1 and 16 gradient accumulation steps. The phoneme classification head uses a learning rate of 5e--3, while the Whisper encoder/decoder is fine-tuned with 10e--6. The chunked self-attention module uses 4 attention heads, a chunk size of 500 frames (10 seconds), and 50\% overlapping chunks. Phonsa is trained for 24k steps based on the validation loss.

For MPEcho, we accordingly curate an internal  Chinese  lyrics-song dataset consisting of both traditional and pop songs. 
We employ SongPrep \cite{tan2025songprep} for structure parsing and Qwen2-Audio-7B \cite{chu2024qwen2} 
for captioning and extracting tags related to genre, instrumentation, and mood/theme. Furthermore, we apply Essentia \cite{bogdanov2013essentia} for key and BPM extraction. These tags are  concatenated to form the input prompts. The dataset comprises 13,045 tracks, totaling $\sim$1,427 hours of audio. We split it into 12,914 tracks for training and 131 tracks for evaluating CSG models.


We train MPEcho and its ablations on a single RTX PRO 6000 with the batch size 1, accumulated to 32. We apply AdamW for optimizing all of the trainable parameters with learning rate 1e--4 and weight decay 0.01, beta (0.8, 0.9). For fair comparison, we train all the variants of our model up to 50k steps. The trainable parameters are 53.3M for IA-EiLM and fusion layers, 330k for the melody encoder, and 12.1M for the phoneme encoder. 

\begin{table}
\centering
\begin{adjustbox}{width=\columnwidth}
\begin{tabular}{l|rcc|cc}
\toprule
\multirow{2}{*}{Model}&\multicolumn{3}{c}{Alignment}&\multicolumn{2}{c}{Segmentation}\\
~ &{MAE}\,$\downarrow$ & {PCO}\,$\uparrow$ & {PCAS}\,$\uparrow$ & {FA}\,$\uparrow$ & {BD F1}\,$\uparrow$ \\
\midrule
MFA~\cite{mcauliffe2017montreal}&233.9\,ms&0.767& 0.680 &-&-\\
\midrule
Phonsa&32.6\,ms& 0.965& 0.897& 0.849& 0.534\\
Phonsa (RNN)&33.2\,ms& 0.963& 0.894&0.794& 0.474\\
Phonsa w/o BT&31.7\,ms& 0.963& 0.899& 0.789& 0.507\\
\bottomrule
\end{tabular}
\end{adjustbox}
\caption{Evaluation of Phonsa variants and MFA for supervised phoneme alignment and unsupervised segmentation.}
\label{tb:phonsa}
\end{table}

\begin{table*}
\centering
\begin{adjustbox}{width=2\columnwidth}
\begin{tabular}{c|ll|cccc|c|cc|c}
\toprule
\multirow{2}{*}{{ID}}&\multirow{2}{*}{{Model}}&\multirow{2}{*}{{Conditions}}&\multicolumn{4}{c}{{Audiobox }} & {Prompt} &\multicolumn{2}{c}{ {Melody}}&  {Lyrics}\\
&  & &  {CE}\,$\uparrow$ &  {CU}\,$\uparrow$ &  {PC}\,$\uparrow$ &  {PQ}\,$\uparrow$ &  {CLAP}\,$\uparrow$ &  {RPA}\,$\uparrow$ &  {RCA}\,$\uparrow$ &  {PER}\,$\downarrow$\\
\midrule
1&SongEcho~\cite{li2026songecho}&\texttt{M}&\underline{6.7389}&\underline{6.9121}&\textbf{6.4657}&\underline{7.4116}&0.1104&0.5779&0.5864&0.4562\\
2&MPEcho (Ours)&\texttt{P}&4.1731&4.5287&5.1739&5.3244&\textbf{0.1576}&0.0667&0.0906&\underline{0.2292}\\
3&MPEcho (Ours)&\texttt{M+P}&\textbf{6.9687}&\textbf{7.0869}&{6.0914}&\textbf{7.5097}&\underline{0.1343}&\underline{0.5764}&\underline{0.5846}&\textbf{0.1865}\\
4&MPEcho (Ours)&\texttt{M+P*}&6.6496&6.8006&\underline{6.3777}&7.1965&0.1230&\textbf{0.6141}&\textbf{0.6217}&0.7125\\
\midrule
&Real songs&---&7.3402&7.5009&6.3953&8.0061&0.0948&---&---&---\\
&ACE-Step~\cite{gong2025ace}&---&7.2349&7.4878&6.2664&7.8369&0.2581&---&---&0.4348\\
\bottomrule
\end{tabular}
\end{adjustbox}
\caption{Objective results of CSG models (IDs 1--4) with different conditioning signals. All methods use the same guidance settings as SongEcho ($\omega=15$, $\alpha=0.5$) for fair comparison. \texttt{M} denotes the melody conditioning, 
\texttt{P} the proposed SVS-style phoneme conditioning inspired by SVS research \cite{ren2019fastspeech},
and \texttt{P*} the alternative Jam-style phoneme conditioning used in JAM~\cite{liu2025jamtinyflowbasedsong} (cf. Section \ref{sec:svs-and-jam}). Best result highlighted and second best underscored.
The last two rows are for reference.}
\label{tb:conditions_comparison}
\end{table*}

\begin{table*}
\centering
\begin{adjustbox}{width=2\columnwidth}
\begin{tabular}{l|ll|cccc|c|cc|c}
\toprule
\multirow{2}{*}{ID} & \multirow{2}{*}{ {Guidance}} &\multirow{2}{*}{ {Hyper Params.}}&\multicolumn{4}{c}{ {Audiobox }} &  {Prompt } &\multicolumn{2}{c}{ {Melody }}&  {Lyrics }\\
& & &  {CE}\,$\uparrow$ &  {CU}\,$\uparrow$ &  {PC}\,$\uparrow$ &  {PQ}\,$\uparrow$ &  {CLAP}\,$\uparrow$ &  {RPA}\,$\uparrow$ &  {RCA}\,$\uparrow$ &  {PER}\,$\downarrow$\\
\midrule
5 & N ($\omega$)&(15.0)&{7.1225}&{7.2432}&5.7535&{7.6736}&0.1546&0.3878&0.3990&0.1780\\
\midrule
6\,($\equiv$3) & \multirow{1}{*}{SE ($\omega$, $\alpha$)}&(15.0, 0.5)&6.9687&7.0869&6.0914&7.5097&0.1343&0.5764&0.5846&0.1865\\
\midrule
7 & \multirow{3}{*}{MC ($\omega_{t}$, $\omega_{l}$, $\omega_{a}$)}&(15.0, 5.0, 2.5)&7.0303&7.1074&5.9021&7.5113&0.1373&0.6121&0.6240&0.2453\\
8 & &(15.0, 7.5, 2.5)&7.0458&7.1221&5.8561&7.5405&0.1332&0.6098&0.6215&0.1865\\
9 & &(15.0, 7.5, 5.0)&7.0161&7.0922&5.7061&7.5268&0.1355&0.6241&0.6344&0.1793\\
\bottomrule
\end{tabular}
\end{adjustbox}
\caption{Comparison of different guidance strategies (cf. Section \ref{sec:cfg}), including na\"ive guidance (`N'), the method used by SongEcho (`SE'), and our multi-condition guidance (`MC'), for the same MPEcho (\texttt{M+P}) model.}
\label{tb:cfg}
\end{table*}

\section{Experiments}

\label{sec:exp_phonsa}
\subsection{Objective Evaluation of Phonsa}

We first evaluate the performance of phonetic transcription tasks, comparing  Phonsa with MFA~\cite{mcauliffe2017montreal} and two ablated versions of Phonsa, one without the boundary tokens (`Phonsa\,w/o\,BT'), the other using the  RNN backbone of \cite{junyou2023transcription} (`Phonsa\,(RNN)') instead of self-attention.

We consider two tasks: alignment and segmentation. 
For \textit{forced alignment}, we adopt three objective metrics. 
Mean absolute error (MAE) measures the average deviation, in milliseconds, between predicted and ground-truth phoneme boundaries. 
Percentage of correct onsets (PCO) computes the proportion of predicted onset boundaries that fall within a tolerance window of 0.1 seconds from the ground truth. 
Percentage of correctly aligned segments (PCAS) evaluates segment-level alignment quality via temporal Intersection over union (IoU) between predicted and ground-truth phoneme segments, normalized by the total ground-truth duration. 
Higher PCO and PCAS, and lower MAE, indicate better alignment performance.
For \textit{unsupervised phoneme segmentation}, we report the frame accuracy (FA), which measures the proportion of correctly predicted phoneme labels at the frame-level. In addition, we adopt boundary F1 score (BD F1), following Strgar \textit{et al.}~\cite{strgar2022phonemesegmentation}, to assess the temporal precision and recall of predicted phoneme boundaries.

Table~\ref{tb:phonsa} shows that Phonsa consistently outperforms the MFA baseline on phoneme-level alignment metrics. In particular, Phonsa greatly reduces the MAE from 233.9 ms to 32.6 ms, indicating a substantial improvement in alignment accuracy. Furthermore, while MFA is limited to forced alignment, Phonsa exhibits broader applicability by supporting unsupervised phoneme segmentation, achieving an FA of 0.849 and a BD F1 of 0.534.
Table~\ref{tb:phonsa} also shows that the ablated versions of Phonsa achieve competitive results for the alignment task but worse results for the segmentation task compared to the proposed default setting.

We observe that boundary detection in singing voice remains challenging, as reflected by the relatively low BD F1 scores under a strict 20 ms tolerance. Nevertheless, Phonsa achieves consistent performance across both alignment and segmentation tasks, suggesting its effectiveness in handling phoneme-level temporal structure for singing.

\subsection{Objective Evaluation of MPEcho}
\label{sec:exp:mpecho}

Next, we move on and report a series of three experiments assessing the effect of different design choices on the performance of MPEcho for CSG, using objective measures. 
For the metrics, we evaluate generation quality from multiple complementary perspectives. 
We use Audiobox~\cite{tjandra2025meta} to assess overall musical quality and aesthetics by content enjoyment (CE), content usefulness (CU), production complexity (PC), and production quality (PQ).
Moreover, we use CLAP~\cite{elizalde2023clap} (the \texttt{music\_audioset\_epoch\_15\_esc\_90.14} checkpoint) to measure the alignment between the generated audio and the text prompt. 
We use \texttt{mir\_eval}~\cite{raffel2014mir_eval} and report raw pitch accuracy (RPA) and raw chroma accuracy (RCA), which measure pitch accuracy with and without octave sensitivity, respectively. 
Finally, we employ SongPrep~\cite{tan2025songprep} for Mandarin lyrics transcription and compute the PER to quantify the phonetic accuracy.


\subsubsection{The Contribution of Phoneme and Melody Controls}

We validate the effectiveness of the proposed phoneme conditioning branch by comparing the following three model variants: 1) SongEcho, which uses the melody conditioning \texttt{M} alone, 2) MPEcho with the proposed phoneme conditioning \texttt{P} alone, 3) the default setting of MPEcho with both the melody and phoneme conditioning (\texttt{M+P}, or `$+$melody$+$phoneme'), i.e., the one depicted in Figure \ref{fig:model}.
For fair comparison, all the three models are implemented by ourselves, trained on the same Chinese Pop dataset, by finetuning the official checkpoint of ACE-Step. Moreover, we fix the CFG strategy to the one used by SongEcho (i.e., following Eq.~(\ref{eq2})), with $\omega=15$ and $\alpha=0.5$.

Table~\ref{tb:conditions_comparison} shows the results of these three models (IDs 1--3), alongside the results of real songs from our test set, and ACE-Step from-scratch generations for references.\footnote{We note that CLAP scores are in general low for all the   methods in Table~\ref{tb:conditions_comparison}. This is likely due to a domain mismatch between the data used to train CLAP and our data distribution, which includes Chinese old songs.}  
We can see that melody conditioning alone (ID 1) yields moderate performance on melody-related metrics (RPA \& RCA), but shows limited improvement in lyrical control, as reflected by a slight increase in PER relative to ACE-Step (0.4348 to 0.4562). 
Using the proposed SVS-style phoneme conditioning alone (ID 2) substantially reduces PER to 0.2292, confirming improved lyric alignment. However, it also causes a marked degradation in Audiobox scores and melody consistency. We attribute this to a fundamental conflict between the high-precision temporal phoneme constraints and the free-form pitch generation of the LTS backbone. Without melody guidance to anchor pitch, the rigid phoneme timing boundaries interfere with the model's natural pitch generation, leading to degraded overall musicality.
Combining melody and SVS-style phoneme conditionings (ID 3) resolves this tension, yielding a more balanced results across RPA, RCA, and PER, effectively reaching the lowest PER 0.1865. This suggests that melody and phoneme conditionings jointly govern musical structure and linguistic content, and that neither alone is sufficient for high-quality result.

\subsubsection{The Effect of Phoneme Temporal Representations}


Next, we compare the SVS-style and JAM-style 
phoneme representations discussed in Section \ref{sec:svs-and-jam}, which differ mainly in their phoneme temporal precision. 
Table~\ref{tb:conditions_comparison} shows that, using the JAM-style conditioning (ID 4) achieves competitive Audiobox scores and strong melody alignment, but at the cost of much higher PER (0.7125), indicating the ineffectiveness of the JAM-style 
representation for fine-grained phoneme control.
We attribute this counter-intuitive result to two compounding factors. First, the Jam-style conditioning scheme inserts multiple vocal filter tokens between phoneme tokens, diluting the conditioning signal and providing ambiguous guidance that degrades vocal quality and subsequently increases transcription errors. 
Second, while such soft constraints may benefit models trained from scratch by allowing implicit learning of intra-word phoneme structure, our setting builds upon a pretrained LTS backbone with a lightweight adapter. We hypothesize that the limited parameter budget of the adapter makes it particularly sensitive to conditioning clarity: ambiguous Jam-style signals are insufficient for the adapter to establish reliable phoneme-to-audio correspondence within the limited training iterations. 

In contrast, the  SVS-style phoneme conditioning (ID 3) provides unambiguous, contiguous phoneme-level boundaries that offer a cleaner gradient signal for adapter optimization, achieving a low PER while maintaining competitive performance across other metrics. This highlights that in adapter-based fine-tuning of pretrained generative models, not only the granularity but also the structural clarity of the conditioning signal is critical for reliable lyric control.

\subsubsection{The Effect of Multi-condition Guidance Strategies} 

Table~\ref{tb:cfg} compares the guidance strategies presented in Section \ref{sec:cfg} and their hyperparameter settings. 
Na\"ive guidance (`N'; ID 5), which applies a single guidance scale (cf.\,Eq.~(\ref{eq1})), yields competitive PER but fails to maintain melody alignment.
The SongEcho guidance method (`SE'; ID 6, equal to ID 3 in Table \ref{tb:conditions_comparison}), which uses two guidance scales (cf.\,Eq.~(\ref{eq2})),  improves both  perceptual quality and melody alignment. 
Lastly, multi-condition guidance (`MC'; IDs 7--9), using three guidance scales (cf.\,Eq.~(\ref{eq3})), further improves performance by explicitly disentangling guidance across modalities, achieving higher Audiobox scores and stronger RPA and RCA compared to SE. While there are trade-offs between Audiobox and PER across MC configurations (IDs 7--9), 
we select `MC~(15.0,~7.5,~5.0)' (i.e., ID 9) as the representative configuration for subjective evaluation reported below, as it achieves the highest RPA, RCA and the lowest PER among the three.

\subsection{Subjective Evaluation of MPEcho}

Finally, we perform a listening test on a set of LLM-generated prompts with melody extracted from real songs to simulate the cover song generation scenario.
A total of 12 audio samples were prepared. Each questionnaire of the test has four questions, each representing a unique text prompt and real lyrics and reference audio, along with four generated audio from different models (randomized and anonymized).
Participants rate each generated audio on a 5-point Likert scale (1$=$poor, 5$=$excellent) across the four dimensions:
\textbf{Melody Consistency (MC)} measures the extent to which the generated audio follows the melody of the reference, including pitch \& temporal alignment.
\textbf{Vocal Naturalness (VN)} reflects the overall robotic timbral artifacts, lyrical mispronunciation, and unnatural prosody found in AI-generated singing.
\textbf{Prompt Adherence (PA)} measures how faithfully the generated audio reflects the input text prompt.
\textbf{Overall (OA)} jointly considers the musical coherence, vocal quality, and overall perceptual quality.

We evaluate the models corresponding to IDs 1, 2, 3, 9 from Tables \ref{tb:conditions_comparison} and \ref{tb:cfg}. Table~\ref{tb:subjective} presents the MOS averaged from 25 participants.
Among IDs 1--3, we can see that using both the melody and phoneme conditioning (i.e., ID 3) leads to consistently higher scores across all the four metrics, while phoneme alone (ID 2) performs the worst. This confirms the complementary benefit of joint melody and phoneme conditioning.
Moreover, Table~\ref{tb:subjective} shows that the MOS can be even higher when choosing the MC~(15.0,~7.5,~5.0) inference strategy (ID 9), leading to the highest MOS across all metrics.
Interestingly, listeners appear to reward both melodic fidelity and lyrical clarity simultaneously in a manner that the objective metrics adopted in Section \ref{sec:exp:mpecho} do not fully reflect, highlighting the importance of subjective evaluation in CSG assessment.

\begin{table}
\centering
\begin{adjustbox}{width=\columnwidth}
\begin{tabular}{c|l|cccc}
\toprule
ID & {Model}&{MC}&{VN}&{PA}&{OA}\\
\midrule
1 & \texttt{M} (SE)&3.36{\small $\pm$1.05}&2.80{\small $\pm$1.03} &2.93{\small $\pm$1.12} & 2.92{\small $\pm$0.99}\\
2 & \texttt{P} (SE)& 1.90{\small $\pm$1.08} & 2.57{\small $\pm$1.06} & 2.71{\small $\pm$1.24} & 2.39{\small $\pm$1.03}\\
3 & \texttt{M+P} (SE)&\underline{3.62}{\small $\pm$1.10} & \underline{3.33}{\small $\pm$0.95} & \underline{3.09}{\small $\pm$1.21} & \underline{3.21}{\small $\pm$0.98}\\
9 & \texttt{M+P} (MC) & \textbf{3.88}{\small $\pm$0.97} & \textbf{3.59}{\small $\pm$1.00} & \textbf{3.37}{\small $\pm$1.14} & \textbf{3.57}{\small $\pm$0.91}\\
\bottomrule
\end{tabular}
\end{adjustbox}
\caption{The subjective evaluation results (mean~$\pm$~standard deviation) comparing different conditioning signals and inference  strategies drawn from Tables \ref{tb:conditions_comparison} and \ref{tb:cfg}.}
\label{tb:subjective}
\end{table}

\section{Conclusion}\label{sec:conclusion}


In this paper, we have presented MPEcho, the first framework integrating explicit phoneme-level condition into LTS to achieve better CSG. By incorporating an SVS-inspired phonetic control branch and a length regulator into melody-guided cover generation, MPEcho provides finer-grained linguistic control and improves lyric intelligibility over prior melody-only approaches. To support scalable phoneme-aware generation, we further introduced Phonsa, an automatic  phoneme alignment system that provides accurate phoneme-level timing and substantially improves temporal alignment quality over conventional forced-alignment baselines. Experimental results show that explicit phoneme-level conditioning complements melody guidance, enabling a more effective balance between melodic consistency and linguistic clarity in controllable cover song generation. These findings highlight the importance of fine-grained phonetic supervision in full-song generation and suggest that structured singing priors from SVS are valuable for improving controllability in end-to-end music generation. 
However, MPEcho is limited to single-singer scenarios. Future work will explore multi-singer generation, multilingual phoneme modeling, and richer prosodic control.

\section{AI Usage Statement}

During the preparation of this work, we used Gemini, ChatGPT to refine the linguistic quality and improve the clarity of the manuscript. After using this service, we reviewed and edited the content carefully and take full responsibility for the content of the published article. The use of these technologies was strictly limited to language editing. The research design, data collection, citation, or technical analysis are all conducted by authors.
\section{Acknowledgment}
The work is supported by grants from Google Asia Pacific, the National Science and Technology Council of Taiwan (NSTC 114-2628-E-002-013-MY3), and the Ministry of Education (MOE) of Taiwan (for Taiwan Centers of Excellence in Artificial Intelligence).
\bibliography{ISMIRtemplate}
\end{document}